\documentclass[runningheads]{llncs}
\usepackage{graphicx}
\usepackage{todonotes}
\usepackage{comment}
\usepackage{float}
\usepackage{indentfirst}
\usepackage{color, colortbl}
\usepackage{wrapfig}
\definecolor{rowHeader}{gray}{0.9}
\usepackage{booktabs} 
\usepackage[english]{babel}
\usepackage[export]{adjustbox}
\usepackage{mathrsfs}
\usepackage{csquotes}
\usepackage{multirow}
\usepackage{multicol}
\usepackage{array}
\usepackage{arydshln}
\usepackage{booktabs}
\usepackage{subfig}
\usepackage{siunitx}
\usepackage[inline]{enumitem}
\usepackage{xspace}
\usepackage{etoolbox}
\usepackage[skip=2pt]{caption}
\DeclareCaptionFont{xbf}{\bfseries\boldmath}
\usepackage{pbox}
\usepackage{siunitx}
\usepackage{amsmath}
\usepackage[square, sort, numbers]{natbib}

\begin{document}
\title{Towards Building Economic Models of Conversational Search}
\author{
Leif Azzopardi\inst{1} \and
Mohammad Aliannejadi\inst{2} \and Evangelos Kanoulas\inst{2}
}
\authorrunning{Azzopardi et al.}

\institute{
University of Strathclyde, Glasgow, UK \and University of Amsterdam, Amsterdam, NL\\
\email{leif.azzopardi@strath.ac.uk,\{m.aliannejadi,e.kanoulas\}@uva.nl}
}

\maketitle              %
\begin{abstract}
Various conceptual and descriptive models of conversational search have been proposed in the literature -- while useful, they do not provide insights into how interaction between the agent and user would change in response to the costs and benefits of the different interactions. In this paper, we develop two economic models of conversational search based on patterns previously observed during conversational search sessions, which we refer to as: \textit{Feedback First} where the agent asks clarifying questions then presents results, and \textit{Feedback After} where the agent presents results, and then asks follow up questions. Our models show that the amount of feedback given/requested depends on its efficiency at improving the initial or subsequent query and the relative cost of providing said feedback. This theoretical framework for conversational search provides a number of insights that can be used to guide and inform the development of conversational search agents. However, empirical work is needed to estimate the parameters in order to make predictions specific to a given conversational search setting.
\end{abstract}
\vspace{-2mm}
\section{Introduction}
\vspace{-1mm}
Conversational Search is an emerging area of research that aims to couch the information seeking process within a conversational format~\cite{DBLP:journals/dagstuhl-reports/AnandCJSS19,DBLP:journals/sigir/Culpepper0S18} -- whereby the system and the user interact through a dialogue, rather than the traditional query-response paradigm~\cite{DBLP:conf/sigir/YanSW16,DBLP:conf/chi/VtyurinaSAC17}. Much like interactive search, conversational search presents the opportunity for the system to ask for feedback, either through clarifying questions or follow up questions in order to refine or progress the search~\cite{DBLP:journals/jasis/CroftT87,belkin1995cases,allen1999mixed}.  While there have been numerous studies trying to develop methods to improve the query clarifications or follow up questions, and to better rank/select results to present/use during a conversational search (e.g.~\cite{DBLP:conf/acl/DaumeR18,DBLP:conf/sigir/AliannejadiZCC19,DBLP:conf/www/ZamaniDCBL20,DBLP:conf/sigir/KieselBSAH18,DBLP:conf/sigir/HashemiZC20,DBLP:journals/corr/abs-2009-11352}) -- less attention has been paid to modelling conversational search and understanding the trade-offs between querying, assessing, and requesting feedback~\cite{azzopardi2018conceptual}. In this paper, we take as a reference point the work of Vakulenko et al.~\cite{vakulenko2019qrfa} and the work of Azzopardi~\cite{DBLP:conf/sigir/Azzopardi11,DBLP:conf/sigir/Azzopardi14}. The former is an empirically derived model of conversational search, called \textbf{QRFA}, which involves querying ($Q$), receiving/requesting feedback ($RF$) and assessing ($A$), while the latter works are economic models of querying ($Q$) and assessing ($A$). In this paper, we consider two ways in which we can extend the economic model of search to include feedback. And then we use these models to better understand the relationships and trade-offs between querying, giving/requesting feedback and assessing given their relative efficiencies and their relative costs. We do so by analysing what the optimal course of interaction would be given the different models in order to minimise the cost of the conversation while maximising the gain. As a result, this work provides a number of theoretical insights that can be used to guide and inform the development of conversational search agents.

\vspace{-2mm}
\section{Background}
\vspace{-1mm}
In their analysis of conversational search sessions, Vakulenko et al.~\cite{vakulenko2019qrfa} found that two common conversational patterns emerged: 
\begin{enumerate}
\item the user issues a  (Q), the system would respond by requesting feedback (RF), the user would provide the said feedback, and the system would either continue to ask further rounds of feedback (RF), or present the results where user assess items (A), and then repeats the process by issuing a new query (or stops); or,
\item the user issues a  (Q), the system would present results where the user assesses items (A), and then, the system requests feedback given the results (RF), the user would provide the said feedback, and the system would present more results (A), the user would assess more items (A), and the system would request further feedback (RF) until the user issues a new query (or stops). 
\end{enumerate}
Inspired by our previous work~\cite{DBLP:conf/cikm/AliannejadiAZK021}, we can think of these two conversational patterns as \textbf{Model 1}: Feedback First, and \textbf{Model 2}: Feedback After. There, of course, are many other possible patterns i.e. feedback before and after, or not at all, and combinations of. In this work, we shall focus on modelling these two ``\textit{pure}'' approaches for conversational search. But, before doing so, we first present the original economic model of search.

\textbf{An Economic Model of Search:} In~\cite{DBLP:conf/sigir/Azzopardi11}, Azzopardi proposed a model of search focused on modelling the traditional query response paradigm, where a user issues a query (Q), the system presents a list of results where the user assesses A items. The user continues by issuing a new query (or stops). We can call this \textbf{Model 0}. During the process, the user issues Q queries and is assumed to assess A items per query (on average). The gain that the user received was modelled using a Cobbs-Douglas production function (see Eq.~\ref{eq:original}). The exponents $\alpha$ and $\beta$ denote the relative efficiency of querying and assessing. If $\alpha=\beta=1$ then it suggests that for every item assessed per query, the user would receive one unit of gain (i.e. an ideal system). However, in practice, $\alpha$ and $\beta$ are less than one, and so the more querying or more assessing naturally leads to diminishing returns. That is, assessing one more item is less likely to yield as much gain as the previous item, and similarly issuing another query will retrieve less new information (as the pool of relevant items is being reduced for a given topic). 

\begin{equation}
    g_{0}(Q_0, A_0) = Q_0^\alpha. A_0^\beta     \label{eq:original}
\end{equation}
Given $A_0$ and $Q_0$, a simple cost model was proposed (Eq.~\ref{eq:original_cost}), where the total cost to the user is  proportional to the number of queries issued and the total number of items inspected (i.e. $Q_0 \times A_0$), where the cost per query is $C_q$ and the cost per assessment is $C_a$.
\begin{equation}
    c_{0}(Q_0, A_0) = Q_0.C_q + Q_0.A_0.C_a\label{eq:original_cost}
\end{equation}
By framing the problem as an optimisation problem~\cite{DBLP:conf/sigir/Azzopardi14}, where the user wants a certain amount of gain, then the number of assessments per query that would minimise the overall cost to the user is given by:
\begin{equation}
A_{0}^{\star} = \frac{\beta.C_q}{(\alpha-\beta).C_a}
\end{equation}
Here, we can see that as the cost of querying increases, then, on average, users should assess more items. While if the cost of assessing increases, on average, users should assess fewer items per query. If the relative efficiency of querying ($\alpha$) increases, then users should assess fewer items per query, while if the relative efficiency of assessing ($\beta$) increases then the users should assess more. Moreover, there is a natural trade-off between querying and assessing, such that to obtain a given level of gain $g(Q_0,A_0)=G$ then as the $A_0$ increases, $Q_0$ decreases, and vice versa.
In the following sections, we look at how we can extend this model to also include the two different forms of feedback (first and after).

\vspace{-2mm}
\section{Models}
\vspace{-1mm}
Below we present two possible models for conversational search given the two pure strategies of feedback first and feedback after. In discussing the models, we will employ a technique called \textit{comparatives statics} -- that is we make statements regarding the changes to the outcomes (i.e. how users would change their behaviour) given a change in a particular variable assuming all other variables remain the same.

\vspace{-2mm}
\subsection{An Economic Model of Conversational Search - Feedback First}
Under Model 1 (Feedback First), each round of feedback, aims to improve the initial query -- and thus drive up the efficiency of querying. To model this, we need to introduce a function, called $\Gamma(F_{1})$ where the efficiency of queries is directly related to how many rounds of feedback are given. We assume for simplicity that the relationship is linear, each round of feedback increases the efficiency of querying by a fixed amount, which we denote as $\gamma_1$. We can then set: $\Gamma(F_{1}) = \gamma_{1}.F_{1} +\alpha$. If no rounds of feedback are given/requested, then  $\Gamma(F_{1}) = \alpha$ -- which results in the original gain function. The gain function under feedback first is:
\begin{equation}
    \label{eq:gain_model2}
    g_{1}(Q_{1}, F_{1}, A_{1}) =  Q_{1}^{\Gamma(F_{1})}.A_{1}^\beta 
\end{equation}
Since the rounds of feedback are only given per query, then the cost model becomes:
\begin{equation}
    \label{eq:cost_model2}
    c_{1}(Q_{1}, F_{1}, A_{1}) = Q_{1}.C_q + Q_{1}.F_{1}.C_f + Q_{1}.A_{1}.C_a
\end{equation}
Again, we can frame the problem as an optimisation problem, where we would like to minimise the cost given a desired level of gain $G$. Then by using a Lagrangian multiplier, differentiating and solving the equations we are at the following expressions:

\begin{equation}
A_{1}^{\star} = \frac{\beta.C_q + F_{1}.C_f }{(\gamma_{1}.F_{1}+\alpha-\beta).C_a}
\end{equation}

\begin{equation}
F_{1}^{\star} = \frac{\beta.C_q + (\alpha-\beta) A_{1}.C_a }{\gamma_{1}.A_{1}.C_a+\beta.C_f}\label{eq_f1}
\end{equation}

From these expressions, we can see that there is a dependency between the two main actions of assessing and providing feedback (unfortunately, we could not reduce down the expression any further). With respect to the optimal number of assessments, we can see that when $F=0$ the model reverts back to the original model. And similarly, if $C_q$ increases, then $A_{1}^{\star}$ increases. If $C_a$ increases,  $A_{1}^{\star}$ decreases. And if $C_f$ increases and $F_1>0$, then $A_{1}^{\star}$ increases. If the number of rounds of feedback increases, then the model stipulates that a user should inspect fewer items (as the query after feedback is more efficient -- and so more effort should be invested into assessing).  For feedback, we see that if $C_q$ or $C_a$ increases, then the user should provide more feedback, or alternatively the system should request more feedback (i.e. $F_{1}^{\star}$ increases). While if the cost of feedback increases then users should provide less feedback (or the system should request less feedback before returning results). Interestingly, if $\gamma_1$ increases, then fewer rounds of feedback are required. This makes sense because if the clarifying question improves the quality of the original query sufficiently, then there is no need for further clarifications of the information need.
More clarifications would only increase the cost of the conversation but not necessarily increase the amount of gain. On the other hand, if $\gamma_1$ is very small, then it suggests that the clarifying questions are not increasing the quality of the query such that it retrieves more relevant material. Thus, asking clarifying questions, in this case, is unlikely to be worthwhile as it will drive up the cost of conversation without increasing the reward by very much.

\vspace{-2mm}
\subsection{An Economic Model of Conversational Search - Feedback After}
Given Model 2 (Feedback After), we can extended the original economic model of search by adding in feedback ($F_{2}$), such that a user issues a query, assess $A_{2}$ items, then provides feedback to refine the query, followed by assessing another $A$ items. The process of giving feedback and assessing, then repeats this $F_{2}$ times. For this kind of conversational interaction,  we can define the gain to be proportional to the number of queries and the number of rounds of feedback per query and the number of assessments per query or round of feedback:
\begin{equation}
    \label{eq:gain_model1}
    g_{2}(Q_{2}, F_{2}, A_{2}) =  Q_{2}^\alpha. (1+F_{2})^{\gamma_{2}}. A_{2}^\beta 
\end{equation}
As before, $\alpha$ and $\beta$ denote the efficiency of querying and assessing, while $\gamma_{2}$ expresses the efficiency of providing feedback. We can see that if $F=0$ (e.g. no feedback), then the model reverts back to the original model in Eq.~\ref{eq:original}. 
\begin{equation}
    \label{eq:cost_model1}
    c_{2}(Q_2, F_{2}, A_2) = Q_{2}.C_q + Q_{2}.F_{2}.C_f + Q_{2}.(1+F_{2}).A_{2}.C_a
\end{equation}

The cost function is also extended to include the rounds of feedback, and the additional assessments per feedback round, where if $F_{2}=0$ then the cost model reverts back to the original cost model. The amount of feedback per query will depend on the cost of the feedback, and its relative efficiency ($\gamma_{2}$).
So if $\gamma_{2}=0$ and $F_2>0$ then there is no additional benefit for providing feedback -- only added cost.

To determine the optimal number of assessments and feedback, for a given level of gain $G$ where we want to minimise the total cost to the user given $G$ we formulated the problem as an optimisation problem.
Then by using a Lagrangian multiplier, differentiating and solving the equations we are at the following expressions for the optimal number of assessments ($A^{\star}$), and the optimal number of rounds of feedback ($F_{2}^{\star}$) :

\begin{equation}
A_{2}^{\star} = \frac{\gamma.(C_q+C_f) - \alpha.(1+F_{2}).C_f }{(\alpha-\gamma).(1+F_{2}).C_a}\label{eq_model2_full_A}
\end{equation}

\begin{equation}
F_{2}^{\star} = \frac{(\gamma-\beta).C_q + (\beta-\alpha).C_f }{(\alpha-\gamma).C_f}\label{eq_f2}
\end{equation}

First, we can see that under this model, it is possible to solve the equations fully. While this may seem quite different -- during the intermediate steps the optimal  $A_{2}^{\star}$ was:
\begin{equation}
A_{2}^{\star} = \frac{\beta.(C_q + F_{2}.C_f) }{(\alpha-\beta).(F_{2}+1)C_a}\label{eq_model2_partial_A}
\end{equation}
\noindent where we can see the relationship between assessing and giving feedback. And, if we set $F_{2}=0$, then the model, again, falls back to the original model in Eq.~\ref{eq:original}. Specifically, assuming feedback is to be given/requested (i.e. $F_{2}>0$), then if the cost of performing the  feedback ($C_f$) increases, on average, a user should examine more items per query and per round of feedback. 
If the cost of assessing ($C_a$) increases, then, on average, a user should examine fewer items per query/feedback. 
Intuitively, this makes sense, as it suggests a user should invest more in refining their need to bring back a richer set of results, than inspecting additional results for the current query or round of feedback.
Now, if the relative efficiency of assessing ($\beta_2$) increases, then a user should examine more items per query/feedback. 

In terms of feedback, we can see that as the cost of querying ($C_q$) increases, then it motivates giving more feedback. While if the cost of feedback ($C_f$) increases, it warrants providing less feedback. This is because querying is a natural alternative to providing another round of feedback. We can also see that the relative efficiencies of querying ($\alpha_2$) to feedback ($\gamma_2$) also play a role in determining the optimal amount of feedback -- such that as $\gamma_2$ increases they users should give more feedback, while if $\alpha_2$ increase they should query more.

\vspace{-3mm}
\section{Discussion and Future Work}
\vspace{-2mm}
In this paper, we have proposed two economic models of conversational search that encode the observed conversational patterns of \textit{feedback first} and \textit{feedback after} given the QFRA work. While these models represent two possible conversational strategies, they do, however, provide a number of interesting observations and hypotheses regarding conversational search paradigms and the role of feedback during conversational sessions. We do, of course, acknowledge that in practice conversational sessions are likely to be more varied. Nonetheless, the insights are still applicable.

Firstly, and intuitively, if the cost of giving feedback either before/first or after increases, then the number of rounds of feedback will be fewer. 
While if the cost of querying increases, then it motivates/warrants requesting/giving more feedback. 
The amount of which depends on the relative costs and efficiencies of each action. 
However, a key difference arises between whether feedback is given before (first) or after. Under feedback first, if the relative efficiency $\gamma_{1}$ (e.g.~answering clarifying questions, etc.) increases, then perhaps ironically less feedback is required. This is because the initial query will be enhanced quicker than when $\gamma_{1}$ is low. Moreover, if the relative efficiency of querying is initially high, then it also suggests that little (perhaps even no) feedback would be required because the query is sufficiently good to begin with and clarifications or elaborations will only result in increased costs. These are important points to consider when designing and developing a conversational search system.
As the decision to give/request feedback is decided by both the gains and the costs involved (i.e.~is it economically viable?). And thus both need to be considered when evaluating conversational agents and strategies.

Of course, such discussions are purely theoretical. More analysis is required, both computationally through simulations and empirically through experiments to explore and test these models in practice. With grounded user data, it will be possible to estimate the different parameters of the proposed models -- to see whether they provide a reasonable fit and valid predictions in real settings. It is also worth noting that another limitation of these models is that they model the average conversational search process over a population of user sessions -- rather than an individual's conversational search process. For example, the quality of clarifying questions asked during the feedback first model is likely to vary depending on the question, this, in turn, suggests that each question will result in different improvements to the original query (i.e.~$\gamma_1$ is not fixed, but is drawn from a distribution). However, the model is still informative, because we can consider what would happen for different values of $\gamma_1$ and determine when feedback would be viable, and at what point it would not be. Once estimates of the costs and relative efficiencies are obtained for a given setting, it will also be possible to further reason about how or what in the conversational process needs to be improved. %
Finally, more sophisticated models of conversational search could also be further developed to analyse different possible mixed strategies. However, we leave such the empirical investigations and further modelling for future work.

\vspace{-3mm}
\subsubsection {Acknowledgements:}
This research was supported by
the NWO (No. 016.Vidi.189.039 and No. 314-99-301), and
the Horizon 2020 (No. 814961).

\bibliographystyle{splncs04}
\bibliography{8_References}
\end{document}